# Laser-written integrated photonics in glass-ceramics Zerodur

Jun Guan[1]

*Department of Engineering Science, University of Oxford, Parks Road, Oxford, OX1 3PJ, UK*

ABSTRACT

We demonstrated fabrication of waveguide and directional coupler in bulk glass-ceramics Zerodur, through femtosecond laser direct-writing. Single mode waveguide propagation loss of no more than 1.5 dB/cm at wavelength of 777 nm was achieved. We expect this will lead to the deployment of Zerodur integrated photonics in fields like space-borne astronomy, quantum technology and fundamental physics, for both economic and technical benefits.

## 1. Introduction

Thanks to its extremely low coefficient of thermal expansion (CTE) and high CTE homogeneity, glass-ceramics Zerodur (SCHOTT AG) has been a popular choice of material in many applications where thermal stability is critical, like ground [1,2] and space-borne [3,4] astronomy, ultra-high precision metrology [5,6], sensing [7] and nanometer-precision engineering [8,9]. As Zerodur has higher than 90% 10 mm-thickness internal transmittance over spectral range 600 nm to 2.3 μm [10], we believe that if photonic integrated devices or circuits could be fabricated in bulk Zerodur, it would significantly benefit the fields like space-borne astronomy, quantum technology and fundamental physics. In space-borne astronomy, integrated photonics in Zerodur could be used to at least partially replace the current bulky and heavy-weight optics [3,4], which would not only reduce the weight and footprint of a payload on board but also enhance its stability and robustness. In quantum technology and fundamental physics, the advent of Zerodur integrated photonics would increase the capacity of space-borne quantum information processing [11-13] and facilitate fundamental discoveries [14,15].

Although it has been investigated before [16], to our best knowledge, no integrated photonics has been successfully fabricated in bulk glass ceramics Zerodur so far. Femtosecond laser direct-writing is a versatile integrated photonics fabrication technique that is compatible with different materials [17,18]. In particular, its intrinsic three-dimensional (3D) fabrication and rapid prototyping nature enables the advances in some fields like the photon-based quantum information processing [19-21] and topological photonics [22,23]. In this report, we demonstrate the fabrication of common building blocks of photonic integrated circuit (PIC) - waveguide and directional coupler (DC) in bulk Zerodur glass ceramics.

## 2. Experiment and methods

The writing system comprised a laser with 170 fs pulse duration, 1 MHz repetition rate and 514 nm wavelength, a 0.5 NA objective lens with which the circularly-polarised beam was focused into a Zerodur chip, and a three-axis translation stage (Aerotech ABL10100L/ABL10100L/ANT95-3-V) that was used to transversely scan the chip relative to the focus. The Zerodur chips were 2 mm thick, 20 mm wide and with maximum length of 20 mm. After fabrication, they were polished, then inspected and tested with following methods if needed.

A transmission microscope (Zeiss Axioplan 2) was used to inspect the chips through overhead or end-facet imaging. Projected refractive index change profile of written track $\Delta n(x)$ was obtained as follows:

$$\begin{aligned} \Delta n(x) &= \frac{\phi(x)-\phi(0)}{D_z(x)}, \\ \phi(x) &= -\frac{2\pi n}{\lambda_0}\int\left(\frac{1}{I(x)}\int\frac{dI(x)}{dz}dx\right)dx \end{aligned} \quad (1)$$

$\phi(x)$ is waveguide phase profile projected on X-axis, which was measured with quantitative phase microscopy (QPM) through solving the transport-of-intensity equation [24]. $\phi(0)$ is the substrate phase value. $D_z(x)$ is the waveguide dimension along projection direction - Y-axis, which we measured through third-harmonic generation (THG) microscopy. $I(x)$ represents projected waveguide intensity profile.

The QPM was performed with the QPM module of a customer-built 3D multifunctional adaptive optical microscope (MAOM) that was detailed in Ref. [25]. The other module of the MAOM was a THG microscopy one that was used to probe a waveguide for its 3D geometrical profile or THG image at any location in a chip. In MAOM, the QPM and THG microscopy modules shared the optical path from objective lens, sample to sample stage to probe the same location of waveguide in QPM and THG without the need of reloading and erroneously repositioning. The THG microscopy module of the 3D MAOM consisted of: a Cr:Forsterite laser (pulse duration 65 fs, repetition rate 100 MHz, wavelength of 1235 nm), two-axis galvanometer scanner, a deformable membrane mirror for aberration correction, 1.15 NA water immersion objective lens, piezo z-translation stage. THG emission was collected in the forward direction and detected with a photomultiplier tube. Aberrations were corrected by sequentially adjusting the amplitudes of the Zernike polynomial modes added to the deformable mirror, in order to maximise the total image intensity. A flip mirror was placed immediately before the objective lens to toggle between THG and QPM mode. The flip mirror was flipped down in THG mode; while in QPM mode, it was flipped up and a green LED is inserted between condenser and filter, image was captured with a CCD camera.

Finally, the optical properties of waveguides and DCs like insertion losses and mode profiles were measured through light coupling-in tests. In the test setup, a polarisation-maintaining (PM) fibre (Thorlabs P1-630PM-FC-5) was used to butt-couple 777 nm (measured with Ocean Optics USB2000 spectrometer) linearly polarised laser beam from polarisation-maintaining fibre-coupled laser (Thorlabs S1FC780PM) to a Zerodur chip. The output from the chip was out-coupled with a microscope objective lens and measured with a photodiode power sensor that was preceded by an iris to avoid uncoupled light. The near-field mode profile (NFMP) was recorded with a CCD camera. The polarisation direction of the laser beam coupled into a chip was controlled through a combination of a manual fibre polarisation controller and a polarimeter (Polarization Analyzer SK010PA-VIS, Schäfter + Kirchhoff GmbH).

---
[1] Corresponding author.
E-mail address: jun.guan@fraunhofer.co.uk
Jun Guan is currently with Fraunhofer Centre for Applied Photonics, Fraunhofer UK Research Ltd, 99 George Street, Glasgow, G1 1RD, U.K.

A range of writing parameters and track configurations were explored in a fabrication and test feedback iteration aiming at lower waveguide propagation loss, better mode profile and higher fabrication efficiency.

## 3. Experiment results

First, we tried to fabricate waveguide in single-track structure, but without success and no normal guided light could be detected in the laser beam coupling-in test. We measured their projected refractive index profiles with the MAOM and confirmed that the refractive index change was negative in femtosecond laser written tracks in Zerodur. As an example, the projected refractive index profile of a single track written with pulse energy of 176 nJ and scan speed of 1 mm/s is shown in Fig. 1, in which the maximum value of $D_z(x)$ was used to replace $D_z(x)$ itself in Equation (1) for simplicity.

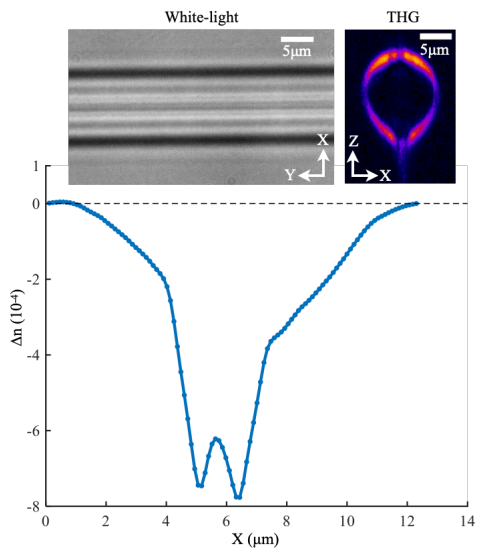

**Fig. 1.** Projected refractive index change profile of a single track written with pulse energy of 176 nJ and scan speed of 1 mm/s. Insets are its overhead white-light microscopy image and THG cross-section microscopy image (in false colour).

Waveguides were successfully fabricated in horizontal double-track configuration. The end-on microscopy image of a such waveguide written with pulse energy of 132 nJ, scan speed of 1 mm/s and track separation of 16 μm is shown in the leftmost inset of Fig. 2. Light was guided between the two tracks. After broadly searching for process window through fabrication and test, we wrote waveguides in different groups to find the one with lowest insertion loss: within each group pulse energy was varied from 132 nJ (at sample level) to 176 nJ; from group to group, scan speed was varied from 0.1 to 6 mm/s and track separation from 8 to 20 μm. These waveguides had the same length of 16.2 mm after polishing. The waveguide insertion losses were tested through coupling in vertical (TM) and horizontal (TE) polarised 777 nm laser. It was found out that waveguides written with separation of 16 μm, scan speed of 1 mm/s and pulse energy of 176 nJ showed the lowest vertical polarised insertion loss among those single-mode waveguides at 777 nm wavelength. The measurement results of the waveguide group written with separation of 16 μm and scan speed of 1 mm/s are shown in Fig. 2. Similar to horizontal double-track waveguides written in other materials [26,27], those waveguides in Zerodur also shows relative low insertion loss at vertical polarisation but very high loss in horizontal polarisation. As shown in Fig. 2, the lowest waveguide insertion loss was 2.4 dB, which corresponded to propagation loss of no higher than 1.5 dB/cm for vertical polarised laser beam with wavelength of 777 nm.

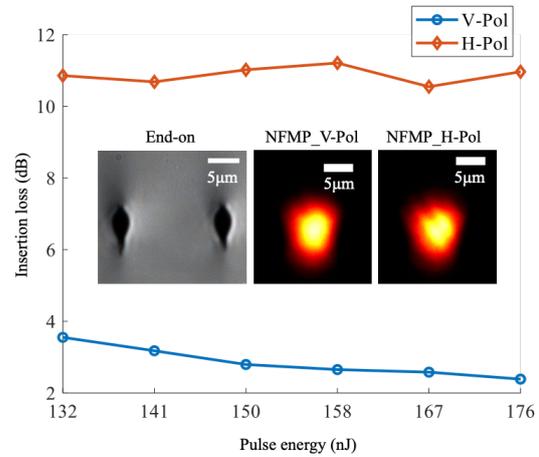

**Fig. 2.** Measured insertion losses of a group of waveguides that were written with scan speed of 1 mm/s, track separation of 16 μm and pulse energy from 132 to 176 nJ. V-Pol and H-Pol denote measurement results with vertical and horizontal polarised 777 nm laser beam respectively. Insets from left to right are end-on microscopy image of a waveguide written at 132 nJ, near-field mode profile (NFMP) images (in false colour) of waveguide written with pulse energy of 176 nJ tested with vertical and horizontal polarised laser beam. The two NFMP images were taken through different neutral density filtering.

Then, we demonstrated directional coupler (DC), a common and versatile building block of photonic integrated circuits. In its simplest form, a DC acts as a beam splitter or combiner through evanescent mode coupling between two waveguides placed in close proximity. The underline principle can be explained with coupled-mode theory [28]. Fig. 3(a). is the schematic of the DCs we fabricated with two horizontal double-track waveguides, in which S is the track separation of a horizontal double-track waveguide; CL is the coupling length; in the coupling region, the two waveguides WR and WL were vertically (Z-axis) aligned with separation of D. Normally the geometrical parameters D or CL will be tuned in order to reach a specific DC beam-splitting ratio. But the waveguide properties in the coupling region will also influence the beam-splitting ratio. In Fig. 3(b), characterisation of two DCs fabricated with following parameters are present as demonstration: pulse energy of 234 nJ, scan speed of 1 mm/s, S = 15 μm, D = 10 μm, CL = 3 or 4 cm, bend radius of 40 mm at all curved regions, 150 μm distance from the top of a DC to the chip top surface. The top figure is end-on microscope image of the DC with CL = 3 cm, the bottom three figures are the NFMP images (in false colour) of the two DCs tested with vertical and 45° polarisation directions. SR is beam splitting ratio. We can see that the DCs were polarisation-sensitive, namely when the polarisation property of coupled-in light changed, the DC exhibited different beam splitting ratio. This is due to the fact that the waveguide effective refractive index and propagation loss are different at different polarisation properties.

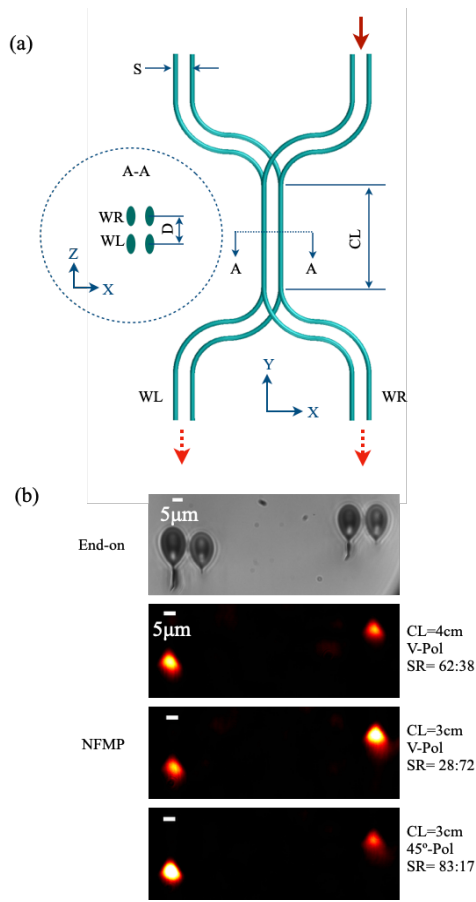

**Fig. 3.** (a) Schematic of the fabricated DCs. S is the track separation of a horizontal double-track waveguide; CL is the coupling length; D is the separation in Z-axis direction between two waveguide WR and WL. (b) Two DCs fabricated with pulse energy of 234 nJ, scan speed of 1 mm/s, S = 15 μm, D = 10 μm and CL = 4 or 3 cm. Top figure is end-on microscope image of the DC with CL = 3 cm; The bottom three figures are NFMP images (in false colour) of the two DCs tested with vertical (V-Pol) and 45º polarisation (45º-Pol); SR is beam splitting ratio.

Other track configurations like vertical double-track, four-track, helical track and double-helical track, as shown in Fig. 4, had also been investigated, but horizontal double track still appeared as the best one in term of waveguide loss, although four-track showed marginal improvement on loss in horizontal polarisation but at the cost of low fabrication efficiency.

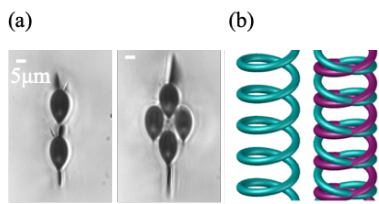

**Fig. 4.** (a) End-on microscopy of a vertical double-track waveguide (left) and a four-track waveguide (right). (b) Schematic of helical and double-helical waveguides.

## 4. Conclusion and outlook

We have successfully demonstrated the fabrication of waveguides and DCs in bulk Zerodur through femtosecond laser direct-writing. As integrated interferometer can be built with waveguides and DCs, we expect this demonstration will lead to the deployment of Zerodur integrated photonics in space-borne astronomy, which will have significant benefits both economically and technically compared to the current payload instruments like interferometers built with standalone heavy-weight bulky optics. Other fields like space-borne quantum science and technology as well as fundamental physics will also benefit from Zerodur integrated photonics in a similar way.

Although the no-higher-than 1.5 db/cm propagation loss waveguide is already directly useful, ever-lower loss waveguide is always demanded in many applications, particularly in photon-based quantum information processing. In addition to lower-loss waveguide, further enrichment of the PIC building blocks like polarisation-insensitive waveguide and DC, waveguides with high extinction ratio in other polarisations and polarisation beam-splitting DCs is also an area of further work. Like laser writing in other materials, the fundamental waveguide formation mechanism is still not clearly known. Even though fundamental mechanisms like stress-induced refractive index change and depressed cladding have been suggested before, any confirmation of these will be helpful for better device fabrication. We note that due to the limitation of our TIE method, we are not able to confirm the existence of stress-induced refractive index change in Zerodur, fused silica or EAGLE 2000 (Corning).

## Declaration of Competing Interest

The authors declare that they have no known competing financial interests or personal relationships that could have appeared to influence the work reported in this paper.

## Acknowledgement

The work was supported by Engineering and Physical Sciences Research Council (EPSRC UK) grant numbers EP/R004803/1 and EP/M013243/1. The work was performed in Dynamic Optics and Photonics group led by Prof. Martin Booth, at University of Oxford.